# Multiphoton Transitions in a Spin System Driven by Strong Bichromatic Field

A. P. Saiko[a], G. G. Fedoruk[b,c], and S. A. Markevich[a]

[a] *Joint Institute of Solid State and Semiconductor Physics, National Academy of Sciences of Belarus, Minsk, 220072 Belarus*
[b] *Institute of Physics, University of Szczecin, 70-451 Szczecin, Poland*
[c] *Sevchenko Institute for Applied Physical Problems, Minsk, 220064 Belarus*
e-mail: saiko@ifttp.bas-net.by

**Abstract**—EPR transient nutation spectroscopy is used to measure the effective field (Rabi frequency) for multiphoton transitions in a two-level spin system bichromatically driven by a transverse microwave (MW) field and a longitudinal radio-frequency (RF) field. The behavior of the effective field amplitude is examined in the case of a relatively strong MW field, when the derivation of the effective Hamiltonian cannot be reduced to first-order perturbation theory in $\omega_1/\omega_{rf}$ ($\omega_1$ is the microwave Rabi frequency, $\omega_{rf}$ is the RF frequency). Experimental results are consistently interpreted by taking into account the contributions of second and third order in $\omega_1/\omega_{rf}$ evaluated by Krylov–Bogolyubov–Mitropolsky averaging. In the case of inhomogeneously broadened EPR line, the third-order correction modifies the nutation frequency, while the second-order correction gives rise to a change in the nutation amplitude due to a Bloch–Siegert shift.

## 1. INTRODUCTION

A subject of current interest is pulsed EPR studies of multiphoton processes in bichromatically driven two-level spin systems [1–6]. The two-level spin system driven by radio-frequency and microwave fields becomes a multilevel system. The significant frequency difference between the MW and RF components of the driving field ensures a high efficiency of multiphoton processes, and the behavior of the dynamic spin system can be controlled by varying the driving field parameters. On the one hand, the photon multiplicity of the processes becomes an adjustable parameter, and the coupling between the microwave field and the spin system can be weakened by varying the RF amplitude so that the medium becomes completely transparent [2]. On the other hand, relaxation can be slowed down by weakening the dipole–dipole coupling and thus increasing the spin decoherence time [7–9]. This offers new possibilities to separate overlapping spectra by enhancing magnetic resonance resolution.

The Rabi frequency associated with multiphoton transitions (henceforth called *multiphoton effective field*) scales linearly with the microwave amplitude and intricately depends on the RF amplitude and frequency. The effective fields calculated by different methods [3, 6, 10] and determined from NMR [10] and EPR [3, 6] data are mutually consistent when the ratio between the microwave Rabi frequency $\omega_1$ and the RF frequency $\omega_{rf}$ is much smaller than unity and calculations can be reduced to first order in $\omega_1/\omega_{rf}$. As the microwave amplitude increases, the condition $\omega_1/\omega_{rf} \ll 1$ is violated and higher order perturbation terms must be taken into account. The multiphoton resonance frequency shifts by an amount of second order in the microwave amplitude (Bloch–Siegert shift), due to nonresonant RF absorption, and the third-order correction to the multiphoton field amplitude becomes important. Calculations of these effects based on different theoretical approaches lead to mathematically different expressions [3, 6]. To the best of our knowledge, strong microwave field effects have never been measured. Therefore, both experimental investigation and adequate theoretical treatment of higher order corrections remain challenging problems of interest not only for EPR and NMR studies, but also for nonlinear optics. This paper presents an experimental and theoretical study of the effective field for multiphoton transitions driven by a strong bichromatic field (with ($\omega_1/\omega_{rf} \leq 0.43$).

## 2. THEORY

The Hamiltonian of the electron (spin-1/2) system interacting with MW and RF fields linearly polarized parallel to the *x* and *z* axes, respectively, in a static magnetic field $B_0$ parallel to the *z* axis can be written as

$$H(t) = \omega_0 s^z + 2\omega_1 \cos(\omega_{mw} t + \varphi) s^x \\ + 2\omega_2 \cos(\omega_{rf} t + \psi) s^z \equiv H_0 + H_1(t) + H_2(t). \quad (1)$$

Here, $\omega_0 = \gamma B_0$ is the resonant transition frequency between spin states; $\gamma$ is the electron gyromagnetic ratio; $\omega_1 = \gamma B_1$ and $\omega_2 = \gamma B_2$ are the Rabi frequencies associated with the driving fields; and $B_1$, $B_2$, $\omega_{mw}$, $\omega_{rf}$, $\varphi$, and $\psi$ are the MW and RF amplitudes, frequencies, and phases, respectively.

When $\omega_1/\omega_{rf}$ or $\omega_2/\omega_{rf}$ is a small parameter, the time evolution of the spin system with Hamiltonian (1) can be described by a perturbation theory using the Krylov–Bogolyubov–Mitropolsky procedure of averaging over fast oscillations [11].

In the canonical Krylov–Bogolyubov–Mitropolsky formalism [11] (see also [12]), fast oscillating terms are eliminated from $H_1(t)$ in each order of perturbation theory in $\omega_1/\omega_{rf}$. The resulting effective Hamiltonian is independent of time, which simplifies analysis of evolution of the dynamical variables of the system. In particular, retaining the third-order terms in the interaction, we have the effective Hamiltonian

$$H_{eff} = H_{eff}^{(1)} + H_{eff}^{(2)} + H_{eff}^{(3)}, \qquad (2)$$

$$H_{eff}^{(1)} = \langle \tilde{H}_1(t) \rangle, \qquad (3)$$

$$H_{eff}^{(2)} = \frac{i}{2} \left\langle \left[ \int^t d\tau (\tilde{H}_1(\tau) - \langle \tilde{H}_1(\tau) \rangle), \tilde{H}_1(t) \right] \right\rangle, \qquad (4)$$

$$H_{eff}^{(3)} = -\frac{1}{3} \left\langle \left\{ \left[ \int^t d\tau (\tilde{H}_1(\tau) - \langle \tilde{H}_1(\tau) \rangle), \left[ \int d\tau \right. \right. \right.\right.$$
$$\left.\left.\left.\left. \times (\tilde{H}_1(\tau) - \langle \tilde{H}_1(\tau) \rangle), \left( \tilde{H}_1(t) + \frac{1}{2}\langle \tilde{H}_1(t) \rangle \right) \right] \right] \right\} \right\rangle \qquad (5)$$

where

$$\tilde{H}_1(t) = \exp(iH_0 t)\exp\left[i\int^t dt' H_2(t')\right]$$
$$\times H_1(t)\exp\left[-i\int^t dt' H_2(t')\right]\exp(-iH_0 t), \qquad (6)$$

$$\langle A(t) \rangle = \frac{1}{T}\int_0^T A(t)dt, \quad T = \frac{2\pi}{\omega_{rf}},$$

and [..., ...] is a commutator.

We calculate $H_{eff}$ under the resonance condition

$$\omega_0 - \omega_{mw} - r\omega_{rf} + \delta = 0, \qquad (7)$$

where $r$ is an integer (positive or negative) number of RF photons (absorbed or emitted, respectively), $\delta$ is an off-resonance detuning ($\delta \ll \omega_{rf}$), and $\omega_0 \gg \omega_{rf}$. Substituting expressions (1) and (6) into (3)–(5), we have

$$H_{eff}(r, \delta t) = H_{eff}^{(1)}(r, \delta t) + H_{eff}^{(2)}(r) + H_{eff}^{(3)}(r, \delta t),$$

where

$$H_{eff}^{(1)}(r, \delta t) = \frac{1}{2}\omega_1^{(1)}(r)e^{-i\delta t}e^{-i(\varphi + r\psi)}s^+ + \text{H.c.}, \qquad (8)$$

$$H_{eff}^{(2)}(r) = \Delta_{BS}(r)s^z, \qquad (9)$$

$$H_{eff}^{(3)}(r, \delta t) = \frac{1}{2}\omega_1^{(3)}(r)e^{-i\delta t}e^{-i(\varphi + r\psi)}s^+ + \text{H.c.}, \qquad (10)$$

$$\omega_1^{(1)}(r) = \omega_1 J_{-r}(z), \qquad (11)$$

$$\omega_1^{(3)}(r) = -\frac{1}{6}\omega_1^3 \left\{ \sum_{n, n' \neq -r} \frac{J_n(z)J_{n'}(z)}{(r+n)\omega_{rf} - \delta} \right.$$
$$\times \frac{J_{n+n'+r}(z) + J_{n'-n-r}(z)}{(r+n')\omega_{rf} - \delta}$$
$$-\frac{1}{2}\sum_{n \neq -r} \frac{J_n(z)J_{-n-2r}(z)J_{-r}(z)}{(r+n)^2 \omega_{rf}^2 - \delta^2} \qquad (12)$$
$$\left. +\frac{1}{2}\sum_{n \neq -r} \frac{J_n^2(z)J_{-r}(z)}{((r+n)\omega_{rf} - \delta)^2} \right\},$$

$$\Delta_{BS}(r) = \frac{1}{2}\sum_{n \neq r} \frac{\omega_1^2}{(r-n)\omega_{rf} - \delta} J_n^2(z), \qquad (13)$$

$s^\pm = s^x \pm is^y$, the argument of the Bessel functions is $z = 2\omega_2/\omega_{rf}$, and the dependence on $\delta t$ in $H_{eff}(r, \delta t)$ is due to the slowly varying factors $\exp(\pm i\delta t)$. These factors are eliminated by changing to a coordinate system rotating with frequency $\delta$ and using condition (7):

$$H_{eff}(r, \delta t) \longrightarrow H_{eff}(r)$$
$$= [\omega_0 + \Delta_{BS}(r) - \omega_{mw} - r\omega_{rf}]s^z + H_{eff}^{(1)}(r) + H_{eff}^{(3)}(r), \qquad (14)$$

where

$$H_{eff}^{(1),(3)}(r) \equiv H_{eff}^{(1),(3)}(r, \delta t = 0).$$

Combining (8), (10), and (14), we find the effective field amplitude to third order in $\omega_1/\omega_{rf}$:

$$\omega_{eff}(r) = \omega_1^{(1)}(r) + \omega_1^{(3)}(r), \qquad (15)$$

where $\omega_1^{(1)}(r)$ and $\omega_1^{(3)}(r)$ are given by expressions (11) and (12).

We note here that the expressions for $\omega_1^{(3)}$ and $\Delta_{BS}$ given, respectively, by Eqs. (18) and (19) in [6] lack a factor of 1/4.

The Hamiltonian $H_{eff}^{(1)}(r)$ describes absorption or emission of a microwave photon and $r$ RF photons. Since the effective bichromatic field amplitude is expressed in terms of Bessel functions of $2\omega_2/\omega_{rf}$, these processes involve multiphoton transitions. For example, the absorption of a microwave photon and $r$ RF photons may involve absorption followed by emission, or emission followed by absorption, of $m$ virtual RF photons, where $m$ is any integer. The contribution $H_{eff}^{(3)}(r)$ of third order in $\omega_1/\omega_{rf}$ to the effective field amplitude is significant when $\omega_1/\omega_{rf}$ is not too small. The Bloch–Siegert shift $\Delta_{BS}(r)$ of a multiphoton resonance frequency in expression (9) for $H_{eff}^{(2)}(r)$ is due to interaction between the spin system and nonresonant RF harmonics. It is clear from (13) that the shift vanishes when $r = 0$ and $\delta = 0$.

The multiphoton nutation driven by a bichromatic field is described by modifying the expression for the one-photon nutation driven by a microwave field in the case of an individual spin packet [13, 14]. In this expression, the microwave Rabi frequency $\omega_1$ should be replaced with the effective bichromatic field amplitude $\omega_{eff}(r)$ given by (15). Furthermore, the off-resonance detuning should include the Bloch–Siegert shift and detuning fluctuations. When $\omega_1 \gg 1/T_2 \gg 1/T_1$, the resulting expression for the multiphoton nutation signal is

$$v(t) \propto \frac{\omega_{eff}(r)}{\Omega(\Delta, r)} \exp\left\{-\frac{1}{T_2}\left(1 - \frac{1}{2}\frac{\omega_{eff}^2(r)}{\Omega^2(\Delta, r)}\right)t\right\} \quad (16)$$
$$\times \sin(\Omega(\Delta, r)t),$$

where

$$\Omega(\Delta, r) = \sqrt{\omega_{eff}^2(r) + \delta^2(\Delta, r)};$$
$$\delta(\Delta, r) = -\delta + \Delta_{BS}(r) + \Delta; \quad (17)$$

$\delta(\Delta, r)$ is the sum of an offset from the line center, a fluctuating detuning $\Delta$, and the Bloch–Siegert shift $\Delta_{BS}(r)$; $T_2$ is the spin–spin relaxation time; $T_1$ is the spin–lattice relaxation time; and $\Omega$ is the nutation frequency. In the case of slow fluctuations, $\Delta$ represents inhomogeneous broadening of the spin transition.

Nutation signal (16) should be averaged over $\Delta$ with the Gaussian weight $g(\Delta) = (T_2^*/\sqrt{\pi})\exp(-\Delta^2 T_2^{*2})$:

$$\langle v(t) \rangle = \int_{-\infty}^{\infty} g(\Delta) v(t) d\Delta, \quad (18)$$

where $T_2^*$ is a reversible dephasing time. Note that the dependence of expressions (11)–(13) on $\Delta$ (via $\delta$) can be neglected.

Integral (18) can be evaluated analytically only when the inhomogeneously broadened linewidth is sufficiently large that $1/\pi T_2^* \gg \omega_1$:

$$\langle v(t) \rangle \propto \omega_{eff} f(\omega_{mw}) J_0(\omega_{eff} t) \exp\left(-\frac{t}{2T_2}\right), \quad (19)$$

where $f(\omega_{mw})$ is the normalized lineshape near the resonance center frequency. When $1/\pi T_2^* \sim \omega_1$, numerical integration is required.

## 3. EXPERIMENTAL

EPR transient nutation spectroscopy was conducted using the 3-cm pulsed EPR spectrometer described in [4]. Resonant coupling between cw microwave and RF fields and the spin system was effected by applying a pulsed longitudinal magnetic field (electron-Zeeman-resolved EPR [4, 6]). The signal-to-noise ratio was improved by multichannel digital averaging. The RF field was not phase-locked to the magnetic field pulse.

The microwave Rabi frequency $\omega_1$ was evaluated by measuring the nutation frequency of the EPR absorption signal detected by setting $\omega_{mw} = \omega_0$ and switching off the RF field. The Rabi frequency $\omega_2$ of the RF field was calibrated by measuring the nutation frequency for dressed spin states at $\omega_{mw} = \omega_0$ and $\omega_{rf} = \omega_1$ [9]. Direct measurements of $\omega_1$ and $\omega_2$ performed under the experimental conditions described below made it possible to evaluate (to within 2%) the parameters required for quantitative comparison with theory and characterization of strong microwave field effects.

The experiments were conducted at room temperature on $E_1'$ centers in neutron-irradiated crystalline quartz (spin-1/2 system). The static magnetic field was parallel to the optical axis of the crystal. For this orientation, the EPR spectrum consists of a single line of width $\Delta B_{pp} = 0.016$ mT. The magnetic field pulse duration was 12 μs, the pulse amplitude was $\Delta B = 0.12$ mT, and the pulse repetition period was 1.25 ms.

## 4. RESULTS AND DISCUSSION

Figure 1 shows typical multiphoton nutation EPR signals from $E_1'$ centers. The corresponding Fourier transforms (insets in Fig. 1) show that the nutation spectrum at low RF field ($z = 0.30$) has a single dominant peak at $\omega_{eff}(0)/2\pi = 0.585$ MHz, which corresponds to multiphoton transitions with $r = 0$. At a higher RF field ($z = 0.97$), the signal spectrum has two dominant peaks at $\omega_{eff}(0)/2\pi = 0.45$ MHz and $\omega_{eff}(1)/2\pi = 0.25$ MHz, which correspond to multiphoton transitions with $r = 0$ and $|r| = 1$. (The dominant nutation frequencies are indicated by arrows in insets.) Several frequencies are simultaneously detected in the nutation spectrum because the RF frequency is comparable to the inhomogeneously broadened EPR linewidth; i.e., when multiphoton transitions with $r = 0$ occur, so do those with $|r| \geq 1$. The dominant contributions to the observed signals come from transitions with $r = 0$ and $\pm 1$. Figure 2 schematizes the transitions with $r = 0$ and 1.

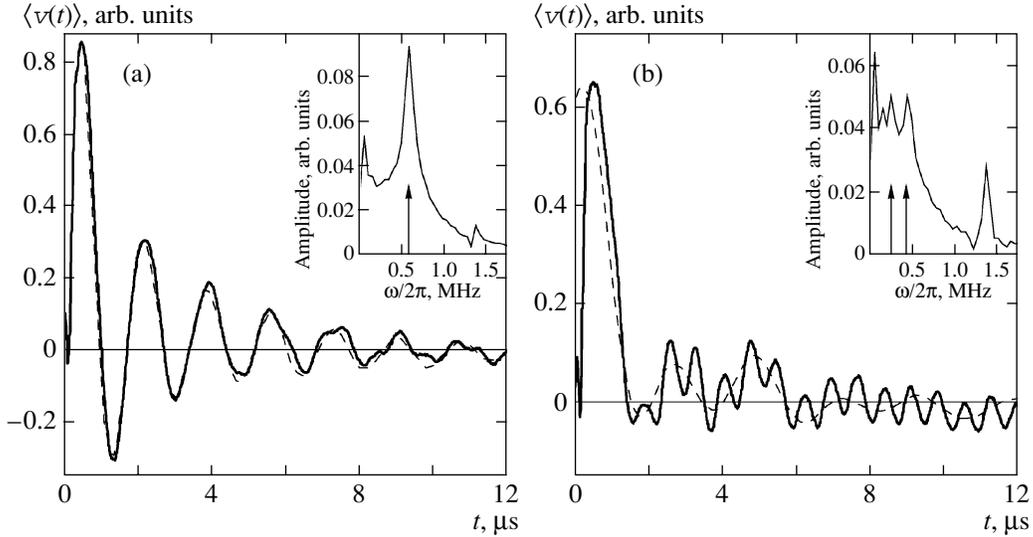

**Fig. 1.** Typical nutation EPR signals from $E'_1$ centers recorded at $\omega_{mw} = \omega_0$, $\omega_1/2\pi = 0.60$ MHz, and $\omega_{rf}/2\pi = 1.38$ MHz. Normalized RF Rabi frequency: (a) $z = 0.30$; (b) 0.97. Dashed curves are approximations. Arrows indicate nutation frequencies corresponding to multiphoton transitions with $r = 0$ (a) and with $r = 0$ and $|r| = 1$ (b).

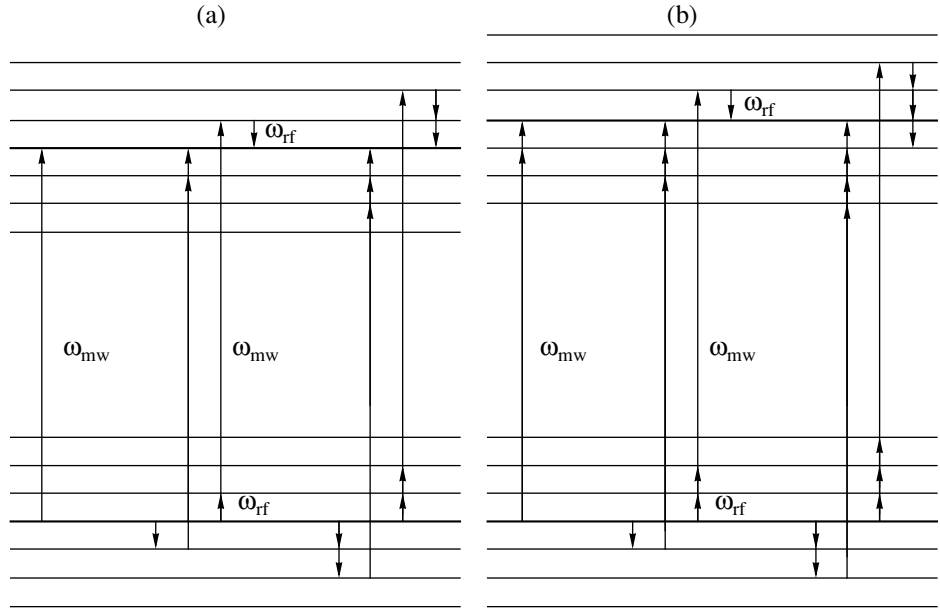

**Fig. 2.** Energy level diagrams for bichromatically driven multiphoton transitions: (a) $r = 0$; (b) $r = 1$. Bold lines represent energy levels of the bare spin system.

Multiphoton transitions with different $r$ are easily identified, because the corresponding nutation frequencies differ as functions of the normalized RF Rabi frequency $z$. Signals due to transitions with $|r| > 1$ are not analyzed here since they are weak and are characterized by low nutation frequencies under our experimental conditions. The RF frequency is much higher, and the corresponding peak in the signal spectrum does not interfere with the detection of nutation frequencies. Dashed curves in Fig. 1 represent the signals approximated by formula (18) for $\omega_{eff}$ specified in the captions, $T_2 = 4$ μs, and $T_2^* = 1$ μs.

Multiphoton nutation frequencies ranging between 0.4 and 0.6 MHz are evaluated to within 2%, and the accuracy decreased with frequency. The domain of Fourier transform is limited from below at a frequency of about 0.15 MHz, determined by the pulse duration. Lower nutation frequencies are evaluated by approximating the observed signals with formula (18).

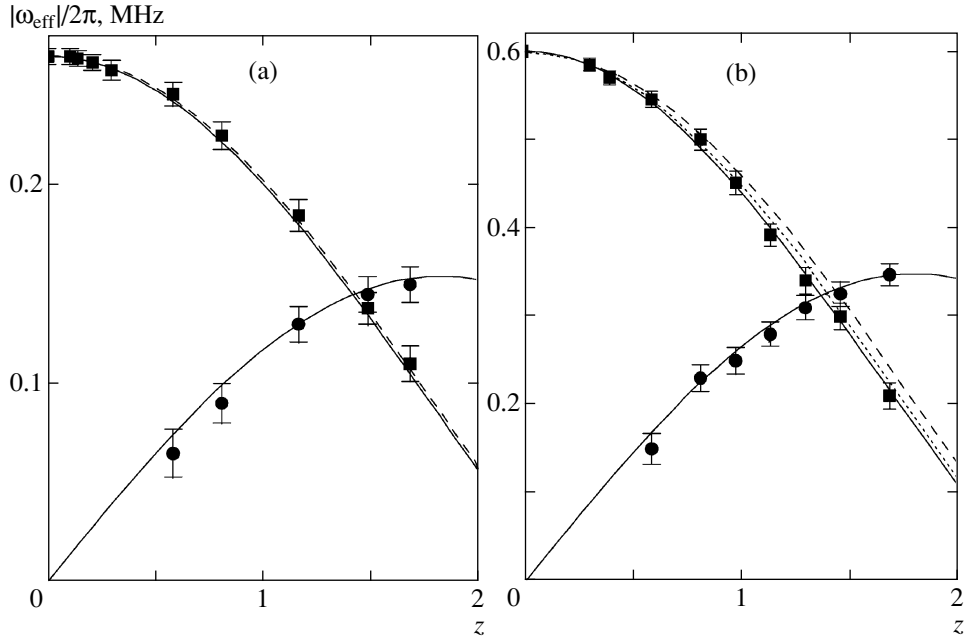

**Fig. 3.** Frequency of bichromatically driven transient nutation vs. normalized RF Rabi frequency $z = 2\omega_2/\omega_{rf}$. Microwave frequency is tuned to the EPR line center; $\omega_{rf}/2\pi = 1.38$ MHz; $\omega_1/2\pi = 0.26$ (a) and 0.60 MHz (b); squares and circles represent experimental data obtained for $r = 0$ and $\pm 1$, respectively. Dashed, solid, and dotted curves are multiphoton effective fields calculated without higher order corrections, with the corrections given by formula (15) in this paper, and with the corrections given by formula (7) in [3], respectively.

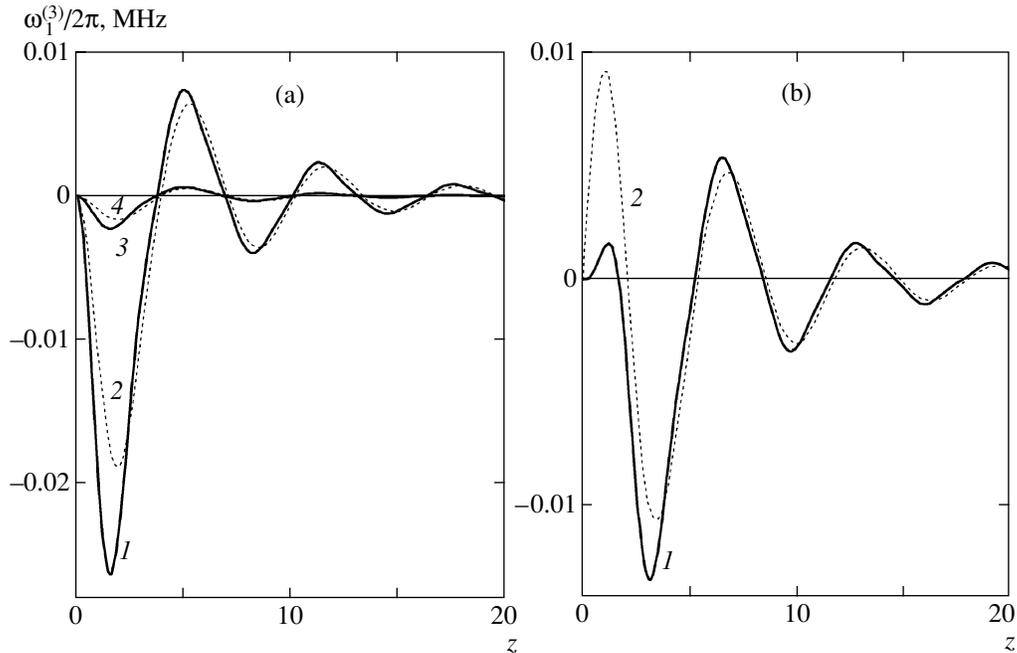

**Fig. 4.** Third-order corrections to multiphoton effective field: (a) $r = 0$ and (b) $r = -1$; $\omega_{rf}/2\pi = 1.38$ MHz; $\omega_1/2\pi = 0.60$ MHz (curves *1* and *2*) and 0.26 MHz (curves *3* and *4*). Solid and dotted curves are calculated by using formula (12) in this paper and formula (7) in [3], respectively.

igure 3 shows the frequencies of bichromatically driven transient nutations as functions of the normalized RF Rabi frequency $z$. Nutation signals were obtained when the spin system was excited by a microwave field tuned to the line center. The RF frequency was held constant at $\omega_{rf}/2\pi = 1.38$ MHz, and the parameter $z$ was varied by varying the RF amplitude. The results presented here were obtained for two values of the MW amplitude (MW Rabi frequency). Increasing $\omega_1/\omega_{rf}$ from 0.19 to 0.43, we determined the third-order contribution to the effective field amplitude for multiphoton transitions with $r = 0$. These transitions are expected to be most strongly modified by the third-order correction $\omega_1^{(3)}(r)$. Because the EPR line is inhomogeneously broadened, the nutation frequency must correspond to $\omega_{eff}(r)$ (see formula (18)). Dashed and solid curves in Fig. 3 represent the multiphoton effective fields evaluated with and without the higher order corrections given by formula (15). The dotted curve in Fig. 3b was calculated by using formula (7) from [3].

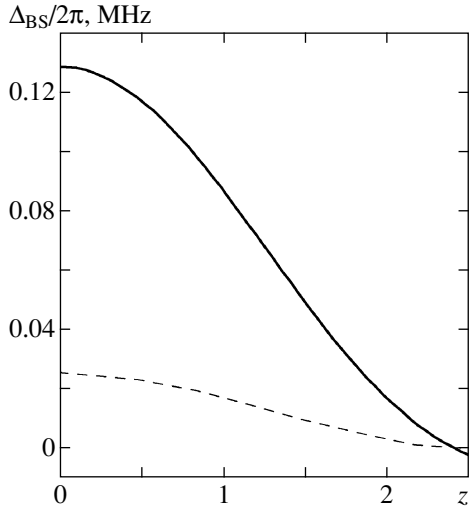

**Fig. 5.** Bloch–Siegert shift calculated for $r = -1$ and $\omega_{rf}/2\pi = 1.38$ MHz. Dashed and solid curves correspond to $\omega_1/2\pi = 0.26$ and 0.60 MHz, respectively.

The results presented in Fig. 3 demonstrate that the contribution of the third-order correction to the effective field amplitude for multiphoton transitions with $r = 0$ is insignificant when $\omega_1/\omega_{rf} = 0.19$ and becomes appreciable as this ratio increases twofold.

Figure 4a compares the third-order contributions to the effective field amplitude for multiphoton transitions with $r = 0$ given by formula (12) above and formula (7) from [3]. It is clear that calculations by two different methods predict qualitatively similar behavior of $\omega_1^{(3)}$ as a function of $z$, while the quantitative difference can amount to 27%.

Because the EPR line is inhomogeneously broadened, the change in the effective field amplitude for multiphoton transitions with $|r| \geq 1$ may have manifested itself via the third-order contribution to the nutation frequency, but the contribution is too small. According to Fig. 3b, the effective field amplitudes for $|r| = 1$ calculated as functions of $z$ with and without the third-order correction are virtually identical. Thus, the third-order effect is significant and experimentally detectable for transitions with $r = 0$.

It should be noted that different theoretical approaches lead to different values of the third-order correction for multiphoton transitions with $|r| \geq 1$ particularly at $z < 2$ (see Fig. 4b). However, the difference is difficult to detect because the correction is small.

Whereas the Bloch–Siegert shift does not occur in multiphoton transitions with $r = 0$, it must modify the nutation frequencies associated with multiphoton transitions with $|r| \geq 1$ via off-resonance detuning. According to (13), the largest Bloch–Siegert shift corresponds to $|r| = 1$. Note that the Bloch–Siegert shift given by expression (6) in [3] obviously exceeds fourfold the correct result.

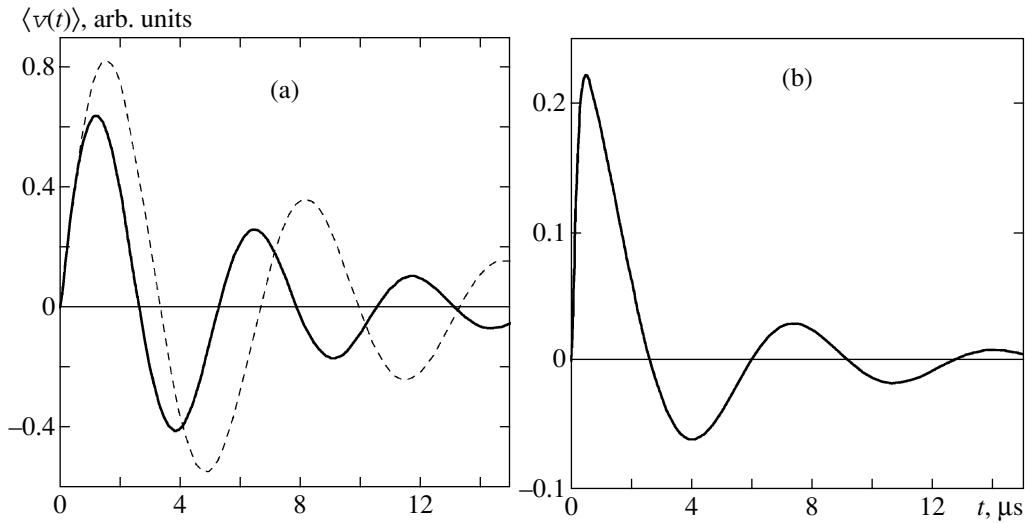

**Fig. 6.** Nutation EPR signal calculated for (a) homogeneously and (b) inhomogeneously broadened EPR line: $r = -1$, $\omega_1/2\pi = 0.60$ MHz, $\omega_{rf}/2\pi = 1.38$ MHz, and $z = 0.52$. Signal without Bloch–Siegert shift, represented by dashed curve, is virtually identical to the actual one in the case of inhomogeneously broadened EPR line (b).

Figure 5 shows the Bloch–Siegert shift versus $z$ for $r = -1$ calculated for the RF and microwave frequencies used in our experiment. Figure 6a demonstrates that the Bloch–Siegert shift significantly changes the effective field amplitude (Fig. 6a), and can therefore be easily detected, in the case of a homogeneously broadened line. However, no change in nutation frequency due to the Bloch–Siegert shift was observed in our experiment, since the EPR signal from $E$ centers in quartz was inhomogeneously broadened. The shift only changed the nutation amplitude, but these changes (less than 1%) were too small to be detected under our experimental conditions, as illustrated by Fig. 6b. If expression (6) in [3] correctly predicted the Bloch–Siegert shift, then the fourfold larger shift (as compared to that given by (13)) would be detectable experimentally. The lack of any observation of this effect may be interpreted as additional evidence that expression (13) given in this paper is correct.

## 5. CONCLUSIONS

EPR transient nutation spectroscopy is used to determine the multiphoton effective field for a bichromatically driven two-level spin system. The experimental results are correctly modeled by using an effective Hamiltonian derived by Krylov–Bogolyubov–Mitropolsky averaging. It is found that the contributions of second and third order in $\omega_1/\omega_{rf}$ increase with the microwave Rabi frequency $\omega_1$. The contribution of second order in $\omega_1/\omega_{rf}$ gives rise to a Bloch–Siegert shift in the EPR line, which could easily be detected as a change in the nutation frequency in the case of a homogeneously broadened spectral line. However, when the line is inhomogeneously broadened (as in our experiment), the shift changes only the nutation amplitude and cannot be detected because the effect is too weak. Therefore, the change in nutation frequency observed in the case of inhomogeneously broadened EPR line is entirely due to the third-order effect of a strong field, offering a simple method for detecting and identifying this effect. The expression for the third-order contribution to the multiphoton effective field obtained by the Krylov–Bogolyubov–Mitropolsky method is in an excellent agreement with experimental results.